\documentclass[12pt]{article}
\begin{document}
\newcommand{\beq}{\begin{equation}}
\newcommand{\eeq}{\end{equation}}
\newcommand{\beqa}{\begin{eqnarray}}
\newcommand{\eeqa}{\end{eqnarray}}
\newcommand{\beqar}{\begin{eqnarray*}}
\newcommand{\eeqar}{\end{eqnarray*}}
\newcommand{\al}{\alpha}
\newcommand{\be}{\beta}
\newcommand{\del}{\delta}
\newcommand{\D}{\Delta}
\newcommand{\eps}{\epsilon}
\newcommand{\ga}{\gamma}
\newcommand{\Ga}{\Gamma}
\newcommand{\ka}{\kappa}
\newcommand{\nn}{\nonumber}
\newcommand{\inn}{\!\cdot\!}
\newcommand{\h}{\eta}
\newcommand{\ii}{\iota}
\newcommand{\kk}{\varphi}
\newcommand\F{{}_3F_2}
\newcommand{\la}{\lambda}
\newcommand{\La}{\Lambda}
\newcommand{\na}{\prt}
\newcommand{\Om}{\Omega}
\newcommand{\om}{\omega}
\newcommand{\p}{\phi}
\newcommand{\sig}{\sigma}
\renewcommand{\t}{\theta}
\newcommand{\z}{\zeta}
\newcommand{\ssc}{\scriptscriptstyle}
\newcommand{\eg}{{\it e.g.,}\ }
\newcommand{\ie}{{\it i.e.,}\ }
\newcommand{\labell}[1]{\label{#1}} 
\newcommand{\reef}[1]{(\ref{#1})}
\newcommand\prt{\partial}
\newcommand\veps{\varepsilon}
\newcommand{\pol}{\varepsilon}
\newcommand\vp{\varphi}
\newcommand\ls{\ell_s}
\newcommand\cF{{\cal F}}
\newcommand\cA{{\cal A}}
\newcommand\cS{{\cal S}}
\newcommand\cT{{\cal T}}
\newcommand\cV{{\cal V}}
\newcommand\cL{{\cal L}}
\newcommand\cM{{\cal M}}
\newcommand\cN{{\cal N}}
\newcommand\cG{{\cal G}}
\newcommand\cH{{\cal H}}
\newcommand\cI{{\cal I}}
\newcommand\cJ{{\cal J}}
\newcommand\cl{{\iota}}
\newcommand\cP{{\cal P}}
\newcommand\cQ{{\cal Q}}
\newcommand\cg{{\it g}}
\newcommand\cR{{\cal R}}
\newcommand\cB{{\cal B}}
\newcommand\cO{{\cal O}}
\newcommand\tcO{{\tilde {{\cal O}}}}
\newcommand\bg{\bar{g}}
\newcommand\bb{\bar{b}}
\newcommand\bH{\bar{H}}
\newcommand\bX{\bar{X}}
\newcommand\bK{\bar{K}}
\newcommand\bA{\bar{A}}
\newcommand\bZ{\bar{Z}}
\newcommand\bxi{\bar{\xi}}
\newcommand\bphi{\bar{\phi}}
\newcommand\bpsi{\bar{\psi}}
\newcommand\bprt{\bar{\prt}}
\newcommand\bet{\bar{\eta}}
\newcommand\btau{\bar{\tau}}
\newcommand\bnabla{\bar{\nabla}}
\newcommand\hF{\hat{F}}
\newcommand\hA{\hat{A}}
\newcommand\hT{\hat{T}}
\newcommand\htau{\hat{\tau}}
\newcommand\hD{\hat{D}}
\newcommand\hf{\hat{f}}
\newcommand\hg{\hat{g}}
\newcommand\hp{\hat{\phi}}
\newcommand\hi{\hat{i}}
\newcommand\ha{\hat{a}}
\newcommand\hb{\hat{b}}
\newcommand\hQ{\hat{Q}}
\newcommand\hP{\hat{\Phi}}
\newcommand\hS{\hat{S}}
\newcommand\hX{\hat{X}}
\newcommand\tL{\tilde{\cal L}}
\newcommand\hL{\hat{\cal L}}
\newcommand\tG{{\widetilde G}}
\newcommand\tg{{\widetilde g}}
\newcommand\tphi{{\widetilde \phi}}
\newcommand\tPhi{{\widetilde \Phi}}
\newcommand\td{{\tilde d}}
\newcommand\tk{{\tilde k}}
\newcommand\tf{{\tilde f}}
\newcommand\ta{{\tilde a}}
\newcommand\tb{{\tilde b}}
\newcommand\tc{{\tilde c}}
\newcommand\tR{{\tilde R}}
\newcommand\teta{{\tilde \eta}}
\newcommand\tF{{\widetilde F}}
\newcommand\tK{{\widetilde K}}
\newcommand\tE{{\widetilde E}}
\newcommand\tpsi{{\tilde \psi}}
\newcommand\tX{{\widetilde X}}
\newcommand\tD{{\widetilde D}}
\newcommand\tO{{\widetilde O}}
\newcommand\tS{{\tilde S}}
\newcommand\tB{{\widetilde B}}
\newcommand\tA{{\widetilde A}}
\newcommand\tT{{\widetilde T}}
\newcommand\tC{{\widetilde C}}
\newcommand\tV{{\widetilde V}}
\newcommand\thF{{\widetilde {\hat {F}}}}
\newcommand\Tr{{\rm Tr}}
\newcommand\tr{{\rm tr}}
\newcommand\STr{{\rm STr}}
\newcommand\hR{\hat{R}}
\newcommand\M[2]{M^{#1}{}_{#2}}

\newcommand\bS{\textbf{ S}}
\newcommand\bI{\textbf{ I}}
\newcommand\bJ{\textbf{ J}}

\begin{titlepage}
\begin{center}

\vskip 2 cm
{\LARGE \bf  Surface terms in the effective actions  \\  \vskip 0.75  cm via duality constraints
 }\\
\vskip 1.25 cm
   Mohammad R. Garousi\footnote{garousi@um.ac.ir}

\vskip 1 cm
{{\it Department of Physics, Faculty of Science, Ferdowsi University of Mashhad\\}{\it P.O. Box 1436, Mashhad, Iran}\\}
\vskip .1 cm
 \end{center}

\begin{abstract}
The effective action of   string theory on a spacetime manifold with boundary has both bulk and boundary terms.   We propose that   both   bulk and  boundary  actions,  may be found by imposing the effective action to be invariant under  the  gauge transformations    and under the string dualities. Using this proposal at the leading order of $\alpha'$, the standard Gibbons-Hawking-York boundary term  is reproduced.
  
\end{abstract}
\end{titlepage}

 String theory is a quantum theory of gravity   with  a finite number of massless fields and a  tower of infinite number of  massive fields reflecting the stringy nature of the gravity. This theory on the spacetime manifolds with boundary is conjectured to be dual to a  gauge theory on the boundary \cite{Maldacena:1997re}. The string theory is usually explored by studying its effective action  which includes the massless fields and their higher derivative terms. For the spacetime manifolds with boundary, the effective action has both bulk and boundary terms, \ie $\bS_{\rm eff}+\prt\!\! \bS_{\rm eff}$. At the leading order of the derivative, the bulk action should  include the Einstein action and   the boundary action should include the Gibbons-Hawking-York term \cite{York:1972sj,Gibbons:1976ue}. These terms and their appropriate  higher derivative corrections should be produced by specific techniques in string theory. 
 
 There are various approaches  for calculating the bulk   actions \eg  the S-matrix approach \cite{Scherk:1974mc,Yoneya:1974jg},  the  sigma-model approach \cite{Callan:1985ia,Fradkin:1984pq}, the Double Field Theory     \cite{ Siegel:1993th,Hull:2009mi} and duality  approach  \cite{Ferrara:1989bc,Font:1990gx,Green:1997tv,Green:2016tfs,Garousi:2017fbe,Green:2019rhz}.
In the duality approach, the  consistency of the effective actions with T- and S-duality transformations are imposed to find the higher derivative couplings. 
 In the T-duality approach, in particular,  the  T-duality  \cite{Giveon:1994fu,Alvarez:1994dn} is imposed as a constraint on the reduction of the effective action on a circle which we call it $S_{\rm eff}$. That is,  the  effective action satisfies the following constraint:
  \beqa
 S_{\rm eff}(\psi)-S_{\rm eff}(\psi')&=&{\rm TD} \labell{TS}
 \eeqa
where $\psi$ represents all  massless fields in the base space and $\psi'$ represents  their transformations under the T-duality  transformations which are the Buscher rules \cite{Buscher:1987sk,Buscher:1987qj} and their higher derivative corrections. On the right-hand side, TD represents some total derivative terms  in the base space which become zero using the Stokes's theorem because the base space has no boundary. This approach has been used in \cite{  Garousi:2019wgz,Garousi:2019mca} to find the effective action of string theory at orders $\alpha'^0,\, \alpha',\, \alpha'^2$ in the bosonic string theory on the closed manifolds. In the superstring theory, there are S-duality as well as T-duality. Imposing S-duality as well as T-duality, one may find couplings in the superstring theory \cite{Green:1997tv,Green:2016tfs,Garousi:2017fbe}. In imposing the S-duality constraint, one should first transforms the couplings to the Einstein frame and then enforcing them to be invariant under the S-duality transformations. In transforming the metric from the string frame to the Einstein frame, one finds some total derivative terms that are again ignored for the closed spacetime. 

It is desirable to extend the above techniques such that they would calculate  the boundary action  $\prt\!\! \bS_{\rm eff}$ as well.   In this paper, we are going to illustrate that a simple extension in  the duality approach  \cite{Garousi:2017fbe} enables  one to calculate  both the bulk and the boundary actions.  

When spacetime  has boundary, the base space in the reduction of the spacetime  on a circle,  has  also  boundary. As a result, the total derivative terms on the right-hand side of \reef{TS} does not vanish using the Stokes's theorem. The total derivative terms resulting from the T-duality of the bulk action   should be cancelled by the T-duality of the boundary action.  
Calling    the reduction of the bulk action   on the  circle  $S_{\rm eff}$  and the reduction of  the  boundary  action   on the same circle   $\prt S_{\rm eff}$, then the T-duality constraint on the effective action \reef{TS} is extended as the following:
 \beqa
 S_{\rm eff}(\psi)+\prt S_{\rm eff}(\psi)&=&S_{\rm eff}(\psi')+\prt S_{\rm eff}(\psi')\labell{TT}
 \eeqa
Unlike \reef{TS}, there are no  total derivative terms in  the base space. 
 The above relation  constrains both bulk and boundary actions. 

Similarly the combination of bulk and boundary actions, \ie $\bS_{\rm eff}+\prt\!\! \bS_{\rm eff}$,  should be written in a S-duality invariant form without ignoring any total derivative term in the bulk. In other words, using the Stokes's theorem, the total derivative terms resulting from transforming the string frame metric  to the Einstein frame metric, produce some boundary terms. They should be combined with the  boundary action $\prt\!\! \bS_{\rm eff}$ to be written in a S-duality invariant form. This S-duality also constrains both bulk and boundary actions. 

Using  appropriate gauge transformations corresponding to the massless fields, one may write the most general couplings in $\bS_{\rm eff}$ and  in  $\prt\!\! \bS_{\rm eff}$, up to  Bianchi identities and field redefinitions  \cite{Metsaev:1987zx}. The arbitrary parameters in the gauge invariant couplings may be fixed by imposing the  above duality constraints. We are going to examine this proposal to find the effective action of superstring theory at the leading order of derivatives, \ie $\bS_{0}+\prt\!\! \bS_{0}$ and for NS-NS fields. In particular, we are going to show that the duality constraints are satisfied only when  the Gibbons-Hawking-York term \cite{York:1972sj,Gibbons:1976ue} is included.

We now construct   the most general $D$-dimensional bulk action  and $(D-1)$-dimensional boundary action at the leading order of $\alpha'$  which are invariant under the coordinate transformations and under the standard  gauge   transformation of $B$-field, \ie $B_{\mu\nu}\rightarrow B_{\mu\nu}+\prt_{[\mu}\lambda_{\nu]}$. Using the fact that in the bulk action the total derivative terms can be transformed to the boundary action by the Stokes's theorem, one finds that in the bulk there are only three terms and in the boundary there  are two terms, \ie 
\beqa
\bS_0&=& -\frac{2}{\kappa^2}\int d^Dx e^{-2\Phi}\sqrt{-G}\,  \left(a_1 R + a_2\nabla_{\mu}\Phi \nabla^{\mu}\Phi+a_3 H^2\right)\,.\labell{S0b}\\
\partial\!\!\bS_0&=&-\frac{2}{\kappa^2}\int d^{D-1}y e^{-2\Phi}\sqrt{\pm g}\,  \left(a_4 G^{\mu\nu}K_{\mu\nu}+a_5n^{\mu}\nabla_{\mu}\Phi \right)\labell{S0b1}
\eeqa
where   the three-form $H$ is field strength of the two-form $B$, \ie $H_{\mu\nu\zeta}=\prt_\mu B_{\nu\zeta}+\prt_\zeta B_{\mu\nu}+\prt_\nu B_{\zeta\mu}$,  In the second  equation, the plus (minus) sign in the square root apply for a spacelike (timelike) boundary,    $K_{\mu\nu}$  is the  extrinsic curvature of the boundary   and  $g_{\alpha\beta}$ is induced metric, \ie
\beqa
K_{\mu\nu}&=&\nabla_\mu n_\nu\,=\,\prt_\mu n_\nu-\Gamma^\zeta{}_{\mu\nu}n_\zeta \nn\\
g_{\alpha\beta}&=&\prt_\alpha x^\mu\prt_\beta x^\nu G_{\mu\nu}\,=\,\frac{\prt x^\mu}{\prt y^\alpha}\frac{\prt x^\nu}{\prt y^\beta}G_{\mu\nu}
\eeqa
where the boundary is specified by the  functions  $x^\mu=x^\mu(y^\alpha)$ and $n^\mu$ is the unit vector orthogonal to the boundary. 
Up to this point  the parameters  $a_1,a_2,a_3,a_4, a_5$,   are arbitrary and above actions are valid for any   theory which has massless fields metric, $B$-field and dilaton. For string theory, however, these parameters should be fixed to specific numbers. We are going to  find them  by the T-duality constraint \reef{TT} and the S-duality constraint.

To impose  the  T-duality constraint on the bulk action, we have to consider a background with $U(1)$ isometry . It is convenient to use the following background for  metric, Kalb-Ramond  and dilaton fields:
  \beqa
G_{\mu\nu}=\left(\matrix{\bg_{ab}+e^{\varphi}g_{a }g_{b }& e^{\varphi}g_{a }&\cr e^{\varphi}g_{b }&e^{\varphi}&}\right),\, B_{\mu\nu}= \left(\matrix{\bb_{ab}+\frac{1}{2}b_{a }g_{b }- \frac{1}{2}b_{b }g_{a }&b_{a }\cr - b_{b }&0&}\right),\,  \Phi=\bar{\phi}+\varphi/4\labell{reduc}\eeqa
where $\bg_{ab}, \bb_{ab}, \bar{\phi} $ are the metric,   the   B-field and the dilaton  in the base space, and $g_{a},\, b_{b}$ are two vectors  in this space. Inverse of the above $D$-dimensional metric is 
\beqa
G^{\mu\nu}=\left(\matrix{\bg^{ab} &  -g^{a }&\cr -g^{b }&e^{-\varphi}+g_{c}g^{c}&}\right)\labell{inver}
\eeqa
where $\bg^{ab}$ is the inverse of the base  metric which raises the index of the   vectors. 
 The Buscher rules \cite{Buscher:1987sk,Buscher:1987qj} in this parametrization become the following linear transformations:
\beqa
\varphi'= -\varphi\,,\, g'_{a}= b_{a }\,,\, b'_a = g_{a } \,\,,\,\bg_{ab}'=\bg_{ab} \,\,,\,\bb_{ab}'=\bb_{ab} \,\,,\,  \bar{\phi}'= \bar{\phi}\labell{T2}
\eeqa
There are higher derivative corrections to these transformations \cite{ Kaloper:1997ux,Garousi:2019wgz} in which we are not interested in this paper.

The proposal \reef{TT} at the leading order of $\alpha'$ can be written as  
\beqa  
S_0(\psi)-S_0(\psi')&=&\prt S_0(\psi')-\prt S_0(\psi)\labell{cons}
\eeqa
where $S_0$ is the reduction of the bulk action  \reef{S0b}  and $\prt S_0$ is the reduction of the boundary action \reef{S0b1} on the circle. Since the bulk actions on the left-hand side are $(D-1)$-dimensional and the boundary actions on the right-hand side are $(D-2)$-dimensional, one expects the left-hand side to be zero up to some  boundary terms which should be  cancelled    by the T-duality transformation of the boundary actions on the right-hand side.

Reduction of different scalar terms in $\bS_0$ are the following  (see \eg \cite{Garousi:2019wgz}):
\beqa
e^{-2\Phi}\sqrt{-G}&\Rightarrow&e^{-2\bphi}\sqrt{-\bg}\nonumber\\
R&\Rightarrow&\bar{R}-\bnabla^a\bnabla_a\vp-\frac{1}{2}\bnabla_a\vp \bnabla^a\vp -\frac{1}{4}e^{\vp}V^2 \labell{R}\\
\nabla_{\mu}\Phi\nabla^{\mu}\Phi&\Rightarrow&\bnabla_a\bphi\bnabla^a \bphi+\frac{1}{2}\bnabla_a\bphi\bnabla^a\vp+\frac{1}{16}\bnabla_a\vp\bnabla^a\vp\nn\\
H^2&\Rightarrow&\bH_{abc}\bH^{abc}+3e^{-\vp}W^2 \nn
\eeqa
 where $V_{ab}$ is field strength of the $U(1)$ gauge field $g_{a}$, \ie $V_{ab}=\bnabla_{a}g_{b}-\bnabla_{b}g_{a}$, and $W_{\mu\nu}$ is field strength of the $U(1)$ gauge field $b_{a}$, \ie $W_{ab}=\bnabla_{a}b_{\nu}-\bnabla_{b}b_{a}$. The    three-form $\bH$ is defined as $\bH_{abc}=\tilde{H}_{abc}-g_{a}W_{bc}-g_{c}W_{ab}-g_{b}W_{ca}$ where the three-form  $\tilde{H}$ is field strength of the two-form $\bb_{ab}+\frac{1}{2}b_{a}g_{b}-\frac{1}{2}b_b g_a $ in \reef{reduc}.  The three-form $\bH$ is invariant under the Buscher rules  \reef{T2} and   satisfies an anomalous  Bianchi identity  \cite{Kaloper:1997ux}.
  
Using the reductions in  \reef{R}, one can calculate the reduced action $S_0(\psi)$ and its transformation $S_0(\psi')$ under the Buscher rules \reef{TT}. Their difference then  becomes 
 \beqa
  S_0(\psi)-S_0(\psi')&=&  -\frac{2}{\kappa^2}\int d^{D-1}x \sqrt{-\bar g}e^{-2\bphi} \,  \Big[ a_2\bnabla^a\bphi\bnabla_a\vp-2a_1 \bnabla^a\bnabla_a\vp\nn\\
 &&\qquad\qquad\qquad\qquad\qquad +(\frac{1}{4}a_1+3a_3)(e^{-\vp}W^2 -e^{\vp}V^2 )  \Big]\labell{s0s0p}
 \eeqa
     One can easily observe that for the following relations between the parameters:
 \beqa
 a_3=-\frac{1}{12}a_1&;&a_2=4a_1\labell{a123}
 \eeqa
The right-hand side of \reef{s0s0p} becomes a total derivative  term in the base space, \ie
\beqa
  S_0(\psi)-S_0(\psi')&=&  \frac{4a_1}{\kappa^2}\int d^{D-1}x \sqrt{-\bar g}\bnabla^a(e^{-2\bphi}\bnabla_{a}\vp)
  \eeqa
The  general form of  the Stokes's theorem relating total derivative of a vector $V^A$ in a bulk  to its value at the  boundary  is 
\beqa
\int_M d^dx\sqrt{|G|}\nabla_A V^A&=&\int_{\prt M}d^{d-1}y\sqrt{|g|}n_{A} V^A
\eeqa
where $G_{AB}$ is the bulk metric,  $g_{IJ}$ is   the induced metric and the boundary is specified by the  functions  $x^A=x^A(y^I)$. In above relation,  $n^A$  is  normal vector to the boundary. It is outward-pointing (inward-pointing) is the boundary is spacelike (timelike).

 It is convenient to use  the Gaussian normal coordinates in the Stokes's theorem.   Using  the  normal coordinates $\{z,y^1,\cdots, y^{D-2}\}$ in the base space, \ie  
\beqa
ds^2\,=\,\bg_{ab}dx^a dx^b\,=\,\sigma d^2z+\tg_{\ta\tb}(z,y^\tc)dy^\ta dy^\tb\,\,; \,or,\,\,\,\,
\bg_{ab}=\left(\matrix{\sigma &  0&\cr0&\tg_{\ta\tb}(z,y^\tc)&}\right) 
\eeqa
where $\sigma=\pm 1$, and     specifying  the boundary as   $x^a=(z_*,y^\ta) $  where boundary is at $z=z_*$, one can write the induced metric in the base space as
\beqa
 g_{\ta\tb}&=&\frac{\prt x^{a}}{\prt y^{\ta}}\frac{\prt x^{b}}{\prt y^{\tb}}\bg_{ab}\nn\\
 &=&\frac{\prt z_*}{\prt y^{\ta}}\frac{\prt z_*}{\prt y^{\tb}}\sigma+\frac{\prt y^{\tc}}{\prt y^{\ta}}\frac{\prt y^{\td}}{\prt y^{\tb}}\tg_{\tc\td}\nn\\
 &=&\tg_{\ta\tb}(z_*,y^\tc)
 \eeqa
The total derivative term in the base space then can be written as the following boundary term: 
 \beqa
  S_0(\psi)-S_0(\psi') &=& \pm\frac{4a_1}{\kappa^2}\int d^{D-2}y \sqrt{\pm\tg(z_*, y^\ta)} e^{-2\bphi}\bnabla^{a}\vp n_a\labell{Tb}
  \eeqa
where $n^a=(1,0,\cdots,0)$  is the outward-pointing  unit vector orthogonal to the boundary, the plus (minus) sign is when the boundary  is spacelike (timelike).

We now turn to the T-duality constraint on  the boundary term \reef{S0b1}. In   the $D$-dimensional  Gaussian normal coordinates $\{z,y^1,\cdots, y^{D-1}\}$, the bulk   metric takes the form 
\beqa
ds^2&=&G_{\mu\nu}dx^\mu dx^\nu\,=\,\sigma d^2z+\gamma_{\alpha\beta}(z,y^\delta)dy^\alpha dy^\beta\,\,;\,or,\,\,\,\,
G_{\mu\nu}=\left(\matrix{\sigma &  0&\cr0&\gamma_{\alpha\beta}(z,y^\delta)&}\right) 
\eeqa 
Inverse of this metric is
\beqa
G^{\mu\nu}=\left(\matrix{\sigma &  0&\cr0&\gamma^{\alpha\beta}(z,y^\delta)&}\right) 
\eeqa 
  The unit vector orthogonal to the boundary in the normal coordinates is  $n^\mu=(1,0,\dots, 0)$.  In using the   Stokes's theorem in the base space, we have   specified   the boundary of the base space as $x^a=(z_*,y^\ta)$, hence, the boundary in the original $D$-dimensional spacetime is specified as   $x^\mu=(z_*,y^\delta) $ where $y^\delta=(y^\ta,y)$ and $y$ is  the  circle along which we have used  the T-duality transformation of the bulk action.  The   induced metric in \reef{S0b1} then becomes
\beqa
 g_{\alpha\beta}&=&\frac{\prt x^{\mu}}{\prt y^{\alpha}}\frac{\prt x^{\nu}}{\prt y^{\beta}}G_{\mu\nu}\nn\\
 &=&\gamma_{\alpha\beta}(z_*,y^\delta)
\eeqa
Since one of the   $y^\delta$  directions is the  circle along which we have used  the T-duality transformation in the bulk action,    the   reduction  of  the boundary metric and its inverse are
\beqa
\gamma_{\alpha\beta}=\left(\matrix{\tg_{\ta \tb}+e^{\varphi}g_{\ta }g_{\tb }& e^{\varphi}g_{\ta }&\cr e^{\varphi}g_{\tb }&e^{\varphi}&}\right)\,;\,\gamma^{\alpha\beta}=\left(\matrix{\tg^{\ta\tb} &  -g^{\ta }&\cr -g^{\tb }&e^{-\varphi}+g_{\tc}g^{\tc}&}\right)
\eeqa
The reduction of different terms in the boundary action  \reef{S0b1} then becomes
\beqa
e^{-2\Phi}\sqrt{\pm g}
&\Rightarrow&e^{-2\bphi}\sqrt{\pm \tg}\nn\\
G^{\mu\nu}K_{\mu\nu}
&\Rightarrow & \bg^{ab}\bK_{ab}+\frac{1}{2}n^a\bnabla_a\vp\nn\\
n^{\mu}\nabla_{\mu}\Phi &\Rightarrow &n^a\bnabla_a\bphi+\frac{1}{4}n^a\bnabla_a\vp
\eeqa
where $\bK_{ab}$ is extrinsic curvature of the boundary in the base space, and we have used the fact that in the Gaussian normal coordinate $n^a=(1,0,\dots, 0)$.  The reduction of the boundary action \reef{S0b1} then becomes
\beqa
\prt S_0(\psi)&=& - \frac{2}{\kappa^2}\int d^{D-2}y \sqrt{\pm\tg(z_*,y^\ta)} e^{-2\bphi}\Big[a_4(\bg^{ab}\bK_{ab}+\frac{1}{2}n^a\bnabla_a\vp)+a_5(n^a\bnabla_a\bphi+\frac{1}{4}n^a\bnabla_a\vp)\Big]\nn
  \eeqa
Then under the Buscher rules \reef{T2}, it transforms as 
\beqa
\prt S_0(\psi)-\prt S_0(\psi')&=& - \frac{2}{\kappa^2}\Big[a_4+\frac{1}{2}a_5\Big]\int d^{D-2}y \sqrt{\pm\tg(z_*,y^\ta)} e^{-2\bphi}n^a\bnabla_a\vp\labell{Ts}
  \eeqa
where we have used the fact that $n^a$ is invariant under the T-duality. 

Replacing the T-duality transformations of the bulk action, \ie \reef{Tb},  and the T-duality transformation of the boundary action, \ie \reef{Ts}, into the constraint \reef{cons}, one finds 
\beqa
a_4&=&\pm 2a_1-\frac{1}{2}a_5
\eeqa
  This relation as well as the relations in \reef{a123}   fix the effective action  up to two parameters  $a_1,a_5$, \ie
\beqa
\bS_0+\prt\!\!\bS_0
&=&-\frac{2a_1}{\kappa^2}\Big[ \int d^{D}x \sqrt{-G} e^{-2\Phi} \left(  R + 4\nabla_{a}\Phi \nabla^{a}\Phi-\frac{1}{12}H^2\right)\pm 2\int d^{D-1}y\sqrt{\pm g}  e^{-2\Phi}K\Bigg]\nn\\
&&-\frac{2a_5}{\kappa^2}\int d^{D-1}y e^{-2\Phi}\sqrt{\pm g}\,  \left(-\frac{1}{2} K+n^{\mu}\nabla_{\mu}\Phi \right)\labell{bbaction}
\eeqa
While there is only one bulk action, there are two boundary actions. In the bosonic theory there is no further constrint that should be imposed to fix the parameter $a_5$.  In the superstring theory however there is still another duality which should be imposed. 

We now impose the S-duality constraint for $D=10$ to fix $a_5$.    To show that  the bulk action for type IIB superstring theory can be rewritten in S-duality invariant form, one   should first change the string frame metric to the Einstein frame metric,     \ie $G_{\mu\nu}=e^{\Phi/2}G_{\mu\nu}^E$. Ignoring a total derivative term resulting from this change of the frames, one finds that after including R-R couplings in which we are not interested in this paper, the Einstein frame couplings  can be written in a S-duality invariant from  (see \eg \cite{Becker:2007zj}).
 However, for the spacetime manifolds with boundary, the total derivative term must be cancelled with the corresponding terms in the boundary.  In fact the total derivative term is produced by transforming the scalar curvature in the bulk action   to the Einstein frame, \ie
\beqa
R&\rightarrow&e^{-\Phi/2}\Big(R-\frac{9}{2}\nabla_\mu\nabla^\mu\Phi-\frac{9}{2}\nabla_\mu\Phi\nabla^\mu\Phi\Big)
\eeqa
The second term above, when replacing it into the bulk action  \reef{bbaction},  produces the following total derivative term in the Einstein frame:
\beqa
\frac{9a_1}{\kappa^2}\int d^{10}x\sqrt{-G^E}\nabla_\mu\nabla^\mu\Phi&=&\pm\frac{9a_1}{\kappa^2}\int d^{9}y\sqrt{\pm g^E}\nabla^\mu\Phi n^E_\mu\labell{total}
\eeqa
where on the right-hand side we have used the Stokes's theorem as well. In this equation, $n^E$ is unite vector orthogonal to the surface in the Einstein frame. This Einstein frame boundary term is not invariant under the S-duality.

   To fully study the S-duality of the boundary terms,  we should also transform the boundary terms in \reef{bbaction} to   the Einstein frame.   Different terms in this action transform as the following:
\beqa
\sqrt{\pm g}&\rightarrow&e^{9\Phi/4}\sqrt{\pm g^E}\nn\\
K&\rightarrow&e^{-\Phi/4}(K^E+\frac{9}{4}n_\mu^E\nabla^\mu\Phi)\nn\\
n_\mu\nabla^\mu\Phi&\rightarrow&e^{-\Phi/4}n^E_\mu\nabla^\mu\Phi
\eeqa
where $K^E=\nabla_\mu(n^E)^\mu$  is the trace of the extrinsic curvature of the boundary  in the Einstein  frame, and we have used the fact that the unite vector $n^\mu$ in the string frame and in the Einstein frame should be related as $n^\mu=e^{-\Phi/4}(n^E)^\mu$, because their lengths are one in both frames, \ie $G_{\mu\nu}n^\mu n^\nu=1=G^E_{\mu\nu}(n^E)^\mu(n^E)^\nu$. 

The total boundary terms in the Einstein frame are then
\beqa
\frac{2}{\kappa^2}\int d^{9}y\sqrt{\pm g^E}\Big[(\mp 2a_1+\frac{1}{2}a_5)K^E+\frac{a_5}{8}\nabla^\mu\Phi n^E_\mu\Big]
\eeqa
The first term is invariant under S-duality. On the other hand, it has been observed in \cite{Garousi:2013qka} that the odd number of  dilaton terms in the Einstein frame can not be combined with the corresponding R-R scalar to be written in a  S-duality invariant form. Hence the  S-duality constrains the coeffient of the last term above to be zero, \ie
\beqa
a_5&=&0
\eeqa
Therefore,  the NS-NS part of the low energy effective action of type II string theories on the spacetime manifolds with boundary can be fixed by the gauge transformations and by the string duality, up to an overall factor $a_1$.  To have the standard Einstein term,  this parameter must be $a_1=1$ as well. So the effective action is 
\beqa
\bS_0+\prt\!\!\bS_0=  -\frac{2}{\kappa^2}\int d^{D}x \sqrt{-G} e^{-2\Phi} \left(  R + 4\nabla_{a}\Phi \nabla^{a}\Phi-\frac{1}{12}H^2\right)\mp\frac{4}{\kappa^2}\int d^{D-1}y\sqrt{\pm g}  e^{-2\Phi}K \labell{S0bf}
\eeqa
where $K$ is the trace of the extrinsic curvature.    For zero dilaton and $B$-field, it is the standard action that its  boundary term has been found by  York, Hawking and Gibbons \cite{York:1972sj,Gibbons:1976ue} by other means. It would be interesting to extend the above calculations to the R-R couplings as well as to the higher orders of $\alpha'$.

 \vskip .3 cm
{\bf Acknowledgements}:   This work is supported by Ferdowsi University of Mashhad.


\end{document}